\documentstyle[prc,aps,preprint]{revtex}

\begin{document}
\draft

\title{A nonequilibrium equality for free energy differences}
\author{C. Jarzynski}
\address{Institute for Nuclear Theory, 
         University of Washington \\
         Seattle, WA~~98195\footnote{ 
         Present address:
         Theoretical Astrophysics, T-6, MS B288,
         Los Alamos National Laboratory,
         Los Alamos, NM 87545} \\
         {\tt chrisj@t6-serv.lanl.gov}} 
\date{\today}

\maketitle

\begin{abstract}
An expression is derived for
the classical free energy difference between two configurations 
of a system, in terms of an ensemble of finite-time
measurements of the work performed
in parametrically switching from one configuration
to the other.
Two well-known equilibrium identities emerge as
limiting cases of this result.
\end{abstract}

\pacs{}

Consider a finite classical system in contact with 
a heat reservoir. 
A central concept in thermodynamics is that of the
{\it work} performed on such a system, 
when some external parameters of the system are made to 
change with time. 
(These parameters may represent, for instance,
the strength of an external field, or the
volume of space within which the system is confined,
or, more abstractly, some particle-particle
interactions which are turned on or off during
the course of a molecular dynamics simulation.)
When the parameters are
changed {\it infinitely slowly} along some
path $\gamma$ from an initial point $A$ to a 
final point $B$ in parameter space, then the
total work $W$ performed on the system is equal to
the Helmholtz free energy difference $\Delta F$
between the initial and final configurations\cite{ll}:
$W=\Delta F\equiv F^B-F^A$.
By contrast, when the parameters are switched along 
$\gamma$ at a {\it finite} rate,
then $W$ will depend on the microscopic
initial conditions of the system and reservoir,
and will on average exceed $\Delta F$:
\begin{equation}
\label{eq:average}
\overline W \ge \Delta F.
\end{equation}
Here and in Eq.\ref{eq:iden} below,
the overbar denotes an average over an {\it ensemble} 
of measurements of $W$,
where each measurement is made after first allowing 
the system and reservoir to equilibrate at temperature 
$T$, with the parameters fixed at $A$.
(The path $\gamma$ from $A$ to $B$, and the rate at
which the parameters are switched along this path, 
remain unchanged from one measurement to the next.)
The difference $\overline W-\Delta F$ is just the 
dissipated work, $W_{diss}$, associated with the 
increase of entropy during an irreversible process.

Eq.\ref{eq:average} is an inequality.
By contrast, the new result derived in this paper 
is the following {\it equality}:
\begin{mathletters}
\label{eq:iden}
\begin{equation}
\label{eq:iden_a}
\overline{\exp -\beta W} = \exp -\beta\Delta F,
\end{equation}
or, equivalently,
\begin{equation}
\label{eq:iden_b}
\Delta F =
-\beta^{-1}\ln\,\overline{\exp -\beta W},
\end{equation}
where $\beta \equiv 1/k_B T$.
\end{mathletters}
This result, which allows one to extract
equilibrium information (the free energy difference
$\Delta F$) from the ensemble of {\it non-equilibrium}
(finite-time) measurements described above, is 
independent of both the path $\gamma$ from $A$ to $B$, 
and the rate at which the parameters are switched 
along the path.

Before proceeding with the proof of Eq.\ref{eq:iden},
we establish notation, and
then relate Eq.\ref{eq:iden} to two well-known
equilibrium identities for $\Delta F$.
Since we have fixed our attention on a particular
path $\gamma$ in parameter
space, it will be convenient to henceforth
view the system as parametrized by a single
quantity $\lambda$, which increases from 0 to 1
as we travel from $A$ to $B$ along $\gamma$.
Let ${\bf z}\equiv ({\bf q},{\bf p})$ denote a point 
in the phase space of the
system, and let $H_\lambda({\bf z})$ denote the 
Hamiltonian for the system, parametrized by the
value of $\lambda$.
Next, let $Z_\lambda$
denote the partition function, let
$\langle\cdots\rangle_\lambda$
denote a canonical average, 
and let $F_\lambda = -\beta^{-1}\ln Z_\lambda$ 
denote the free energy, all 
with respect to the Hamiltonian $H_\lambda$
and the temperature $T$.
We are interested in the following scenario,
which we will refer to as ``the switching process'':
the system evolves, in contact with a heat 
reservoir, as the value of $\lambda$ is switched
from 0 to 1, over a total switching time $t_s$.
Without loss of generality, assume a 
constant switching rate, $\dot\lambda=t_s^{-1}$.
For a given realization of the switching process,
the evolution of the system is described
by a (stochastic) trajectory ${\bf z}(t)$, 
and the work performed 
on the system is the time integral of 
$\dot\lambda\,\,\partial H_\lambda/\partial\lambda$ 
along this trajectory:
\begin{equation}
\label{eq:work}
W = \int_0^{t_s} dt\,\dot\lambda\,
{\partial H_\lambda\over\partial\lambda}
\Bigl({\bf z}(t)\Bigr).
\end{equation}
Now imagine an {\it ensemble} of 
realizations of the switching process
(with $\gamma$ and $t_s$ fixed), 
with the microscopic initial conditions for 
the system and reservoir generated
from a thermal equilibrium ensemble at temperature $T$.
Then $W$ may be computed separately for each 
trajectory ${\bf z}(t)$ in the ensemble, and
the overbars appearing in Eqs.\ref{eq:average}
and \ref{eq:iden} indicate an average over the
distribution of values of $W$ thus obtained.

In the limiting cases of infinitely slow and 
infinitely fast switching of the external parameters,
we know explicitly the ensemble distribution of
values of $W$, and thus can readily check the
validity of our central result.
In the slow limit ($t_s\rightarrow\infty$),
the system is in quasi-static
equilibrium with the reservoir throughout the
switching process, hence
$W=\int_0^1 d\lambda\,
\langle
\partial H_\lambda/\partial\lambda\rangle_\lambda$
for every trajectory in the ensemble.
Eq.\ref{eq:iden_b} then reduces to:
\begin{equation}
\label{eq:adia}
\Delta F = \int_0^1 d\lambda\,
\Biggl\langle
{\partial H_\lambda\over\partial\lambda}\Biggr\rangle_\lambda.
\end{equation}
In the opposite limit ($t_s\rightarrow 0$), 
the switching of the Hamiltonian is 
instantaneous, and so the work performed is simply
$W=H_1-H_0\equiv\Delta H$, 
evaluated at the initial conditions\cite{walls}.
Since we have a canonical distribution of initial
conditions, Eq.\ref{eq:iden_b} becomes, in this case:
\begin{equation}
\label{eq:dia}
\Delta F = -\beta^{-1}\ln\,
\Bigl\langle \exp -\beta\Delta H\Bigr\rangle_0.
\end{equation}
These two results, Eqs.\ref{eq:adia} 
and \ref{eq:dia}, are
well-established identities for the free energy
difference $\Delta F$\cite{kirkwood,zwanzig}.
Note that both give $\Delta F$ in terms of
equilibrium (canonical) averages.
By contrast, 
in the intermediate case of finite $t_s$,
our ensemble of trajectories lags 
behind the equilibrium distribution in phase space
as $H_\lambda$ changes with time.
In this sense, Eq.\ref{eq:iden} is 
an explicitly {\it non}-equilibrium result.

To prove our central result, 
it is instructive to first consider what happens when 
there is {\it no} heat reservoir during the
switching process.
The evolution of the system
is then described by a deterministic trajectory ${\bf z}(t)$
which evolves 
under $H_\lambda({\bf z})$,
as $\lambda$ changes from 0 to 1 over
a time $t_s$.
Consider an ensemble of such trajectories,
defined by a canonical distribution of
initial conditions (at a temperature $T$).
This ensemble is described by a phase space
density $f({\bf z},t)$ 
which satisfies $f({\bf z},0)=Z_0^{-1}\exp -\beta H_0({\bf z})$,
and which evolves under the Liouville equation, 
$\partial f/\partial t+
\{f,H_\lambda\} = 0$, with $\lambda=\lambda(t)=t/t_s$.
Here, $\{\cdot,\cdot\}$ denotes the Poisson bracket.
Since the evolution is deterministic, 
a particular trajectory in this ensemble
is uniquely specified by single point:
there is exactly one trajectory which passes through
a given ${\bf z}$ at time $t$.
This means we can define a ``work accumulated''
function $w({\bf z},t)$, as follows.
For the trajectory which passes through the point
${\bf z}$ at time $t$, $w({\bf z},t)$ is the 
work performed on that trajectory (the time
integral of $\dot\lambda\partial H_\lambda/\partial\lambda$)
up to time $t$.
Since the total work $W$ is just the work
accumulated up to time $t_s$ (Eq.\ref{eq:work}),
the ensemble average $\overline{\exp -\beta W}$ 
may be expressed as
\begin{equation}
\label{eq:av1}
\overline{\exp -\beta W} = 
\int d{\bf z}\,
f({\bf z},t_s)\,
\exp -\beta w({\bf z},t_s).
\end{equation}
Now, the work done on an isolated Hamiltonian system 
is equal to the change in its energy.
Thus, $w({\bf z},t) = H_\lambda({\bf z})-H_0({\bf z}_0)$,
where ${\bf z}_0={\bf z}_0({\bf z},t)$ 
is the initial condition for the
trajectory which passes through ${\bf z}$ at time $t$,
and $\lambda=\lambda(t)$.
Furthermore, Liouville's theorem tells us that phase
space density is conserved along any trajectory,
hence
$f({\bf z},t) = f({\bf z}_0,0) = 
Z_0^{-1} \exp -\beta H_0({\bf z}_0)$.
Combining these results immediately gives
\begin{equation}
f({\bf z},t) \exp -\beta w({\bf z},t)
= Z_0^{-1} \exp -\beta H_\lambda({\bf z}).
\end{equation}
Eq.\ref{eq:av1} then becomes
\begin{equation}
\label{eq:ratio}
\overline{\exp -\beta W} = 
Z_0^{-1}\int d{\bf z}\,\exp -\beta H_1({\bf z})
= Z_1/Z_0.
\end{equation}
Since $\Delta F=-\beta^{-1}\ln (Z_1/Z_0)$,
we have established the validity of
Eq.\ref{eq:iden} for the case in which 
the system is isolated during the switching process.

Now consider the situation in which the
system is coupled to a reservoir.
We assume that the system of interest and
the reservoir together constitute a larger,
{\it isolated} Hamiltonian system.
Let ${\bf z}^\prime$ denote a point in the
phase space of the reservoir, let 
${\cal H}({\bf z}^\prime)$ be the Hamiltonian
for the reservoir alone, and
let ${\bf y}=({\bf z},{\bf z}^\prime)$
denote a point in the full phase space of
system and reservoir.
Motion in the full phase space is deterministic,
and governed by a Hamiltonian 
$G_\lambda({\bf y}) = 
H_\lambda({\bf z}) + {\cal H}({\bf z}^\prime) + 
h_{int}({\bf z},{\bf z}^\prime)$,
where the interaction term $h_{int}$ couples the
system of interest to the reservoir.
Let $Y_\lambda$ be the partition function
for $G_\lambda$.
We explicitly assume the reservoir to be large 
enough, and the interaction energy $h_{int}$ small
enough\cite{reif}, that when $\lambda$ is held fixed 
the system of interest samples its phase space 
according to the Boltzmann factor
$e^{-\beta H_\lambda({\bf z})}$.
Now imagine that, at $t=0$, we populate
the {\it full} phase space with a canonical 
distribution of initial conditions, 
using the Boltzmann factor $e^{-\beta G_0({\bf y})}$.
(This corresponds to allowing the coupled
system and reservoir to equilibrate at temperature 
$T$, before each realization of the switching process.)
From this ensemble of initial conditions, 
an ensemble of trajectories ${\bf y}(t)$
evolves deterministically under $G_\lambda$,
as $\lambda$ switches from 0 to 1.
Since the system of interest and reservoir
together constitute an isolated Hamiltonian
system, the work $W$ performed on the system
of interest is equal to the change in the 
{\it total} energy of the system and reservoir:
$W = G_1 \Bigl({\bf y}(t_s)\Bigr) - 
G_0 \Bigl({\bf y}(0)\Bigr)$.
Therefore, applying the analysis of the previous
paragraph to the situation considered here, with 
${\bf y}$, $G_\lambda$, and $Y_\lambda$ 
replacing
${\bf z}$, $H_\lambda$, and $Z_\lambda$,
respectively, we get
\begin{equation}
\label{eq:reserv}
\overline{\exp -\beta W} = Y_1/Y_0.
\end{equation}
The right side of Eq.\ref{eq:reserv} depends only
on the initial and final Hamiltonians 
$G_0$ and $G_1$, and on the temperature $T$,
which means that
the ensemble average $\overline{\exp -\beta W}$
is independent of the switching time $t_s$
(and also of the path from $A$ to
$B$ in parameter space).
But we already know that
$\overline{\exp -\beta W}=\exp -\beta\Delta F$
in the limit $t_s\rightarrow\infty$,
since $W=\Delta F$ for every member of the
ensemble, in that limiting case.
We therefore conclude that 
\begin{equation}
\label{eq:exps}
\overline{\exp -\beta W}=\exp -\beta\Delta F
\end{equation}
for {\it all} values of $t_s$ (and all paths $\gamma$).
This proves our central result, Eq.\ref{eq:iden}. 

Eq.\ref{eq:reserv}, which tells us that the 
ensemble average $\overline{\exp -\beta W}$ is
independent of both $\gamma$ and $t_s$, is 
identically true, given the formulation of 
the problem.
However, in going from Eq.\ref{eq:reserv} to
Eq.\ref{eq:exps}, we invoke a result from 
quasi-equilibrium statistical mechanics,
which relies on the assumption of weak coupling
(small $h_{int}$).
Eq.\ref{eq:iden}, therefore, is valid 
{\it for sufficiently weak coupling} 
between the system of interest and the reservoir.
This may be seen more directly by writing
an explicit expression for the ratio 
$Y_1/Y_0$ appearing on the right side of 
Eq.\ref{eq:reserv}:
only if $h_{int}$ may be
neglected does this ratio immediately reduce
to $Z_1/Z_0$ ($=\exp -\beta\Delta F$).

Note that the inequality $\overline W\ge\Delta F$
(Eq.\ref{eq:average}) follows directly from
the equality 
$\overline{\exp -\beta W}=\exp -\beta\Delta F$
(Eq.\ref{eq:iden_a}),
by application of the mathematical identity 
$\overline{\exp x}\ge\exp\overline{x}$\cite{chandler}.
This establishes $\overline W\ge\Delta F$ directly
from a microscopic, Hamiltonian basis, rather than 
by invoking the increase of entropy.
(In the limit $t_s\rightarrow 0$, 
we have $\overline W = \langle\Delta H\rangle_0$,
and Eq.\ref{eq:average} reduces to the 
Gibbs-Bogoliubov-Feynman bound\cite{chandler},
$\langle\Delta H\rangle_0\ge\Delta F$.)

It is also worthwhile to point out that the right side 
of Eq.\ref{eq:iden_b} may be expanded as a sum
of cumulants (see Eq.[9] of Ref.\cite{zwanzig}):
\begin{equation}
\Delta F = \sum_{n=1}^\infty
(-\beta)^{n-1}{\omega_n\over n!},
\end{equation}
where $\omega_n$ is the $n$'th cumulant 
of the ensemble distribution of values of $W$.
If this distribution happens to be Gaussian
(as may be expected for sufficiently slow switching),
then only the first two terms survive, and
we have
\begin{equation}
\label{eq:gauss}
\Delta F = \overline W - \beta\sigma^2/2,
\end{equation}
where $\sigma^2\equiv\overline{W^2}-\overline W^2$
is the ensemble variance of $W$.
The dissipated work $W_{diss}$
($=\overline W-\Delta F$) is then related to
the fluctuations in $W$ by:
$W_{diss}=\beta\sigma^2/2$.
This is a fluctuation-dissipation relation,
and has been obtained within the context of
numerical simulations by Hermans\cite{hermans}.
(A related result for microcanonical 
ensembles has been derived by Ott\cite{ott}.)

The central result of this paper, Eq.\ref{eq:iden},
makes a concrete prediction regarding
the outcome of an ensemble of measurements, 
which in principle is subject to experimental
verification.
Moreover, this result ought to be valid 
quite generally, provided the coupling to 
the reservoir is sufficiently weak, and 
quantal effects may be ignored. 
In practice, however, the {\it applicability} of
Eq.\ref{eq:iden} may be severely limited by the 
following considerations.
If the fluctuations in $W$ from one
measurement to the next are much larger than
$k_B T$ (i.e.\ if $\sigma\gg\beta^{-1}$), 
then the ensemble average of $\exp -\beta W$
may be dominated by values of $W$ many
standard deviations below $\overline W$.
Since such values of the work
represent statistically very rare events,
it would require an unreasonably large
number of measurements of $W$ to determine
$\overline{\exp -\beta W}$ with good accuracy.
Therefore, given a
specific system of interest, switching path
$\gamma$, and switching time $t_s$, the
fluctuations in the work $W$ must
not be much greater than $k_B T$, if we
are to have any hope of verifying Eq.\ref{eq:iden}
experimentally.
This condition pretty much rules out macroscopic
systems of interest.
In recent years, however, the direct manipulation
of {\it nanoscale} objects --- and the measurement 
of forces thereon\cite{nano} --- has become feasible.
Such systems may offer the best chance for 
experimentally testing the new result of 
this paper.

So far, we have implicitly assumed that our system
is coupled to a {\it physical} heat reservoir.
It is interesting, however, to discuss this 
problem within the context of numerical simulations.
On a computer, a heat reservoir must somehow be
``mocked up''.
One way to accomplish this is with
a Nos\' e-Hoover (NH) thermostat\cite{nh}, 
or some variant thereof.
In its simplest form, this method replaces the reservoir
with a single variable $\zeta$;
motion in the extended phase space $({\bf z},\zeta)$
is governed by the NH equations
\begin{eqnarray}
\label{eq:nh1}
\Bigl\{
\dot q = p/m\,&,&\,
\dot p = -\nabla\Phi_\lambda-\zeta p
\Bigr\}_n\\
\label{eq:nh2}
\dot\zeta &=& 
(K/K_0-1)/\tau^2.
\end{eqnarray}
[We have assumed a kinetic + potential Hamiltonian:
$H_\lambda=p^2/2m+\Phi_\lambda({\bf q})$.
The index $n$ runs over all $D$ degrees of 
freedom of the system, 
$K=p^2/2m$ is the total kinetic energy of the system,
$K_0=\beta^{-1}D/2$ is the thermal average of $K$, and 
$\tau$ is a parameter which acts as a relaxation time.]
For $\lambda$ fixed,
a trajectory ${\bf z}(t)$ generated by these equations
of motion samples phase space according to the 
Boltzmann factor $\exp -\beta H_\lambda({\bf z})$,
provided that the evolution is sufficiently chaotic.

It is interesting to ask, does Eq.\ref{eq:iden}
remain valid if the system evolves under the NH equations,
rather than under the influence of a physical reservoir?
Let us consider an ensemble of initial
conditions in the extended phase space,
described by the density
\begin{equation}
\label{eq:nhic}
f({\bf z},\zeta,0) = 
c Z_0^{-1} \exp -\beta Q_0({\bf z},\zeta),
\end{equation}
where 
$Q_\lambda({\bf z},\zeta)\equiv
H_\lambda({\bf z})+D\zeta^2\tau^2/2\beta$,
and $c=(D\tau^2/2\pi)^{1/2}$ is a normalization factor.
(The distribution
$cZ_\lambda^{-1}\exp-\beta Q_\lambda$ is 
stationary under the NH equations when $\lambda$ 
is held fixed, and may be viewed as the ``canonical'' 
distribution in the extended phase space.)
Allowing these initial conditions to evolve
under the NH equations, as $\lambda$ changes from
0 to 1, we obtain an ensemble of trajectories
described by a time-dependent density 
$f({\bf z},\zeta,t)$.
As before, the work performed on each member
of the ensemble is defined to be the time integral
of $\dot\lambda\,\partial H_\lambda/\partial\lambda$.
We now introduce a ``work accumulated'' function
$w({\bf z},\zeta,t)$, analogous to $w({\bf z},t)$
introduced earlier.
It is straightforward to establish that
\begin{eqnarray}
f({\bf z},\zeta,t) &=&
f({\bf z}_0,\zeta_0,0) 
\exp\,\Bigl[D\int_0^t\zeta(t^\prime) dt^\prime\Bigr]\\
w({\bf z},\zeta,t) &=&
Q_\lambda({\bf z},\zeta)-
Q_0({\bf z}_0,\zeta_0) + 
\beta^{-1}D\int_0^t\zeta(t^\prime) dt^\prime,
\end{eqnarray}
where $({\bf z}_0,\zeta_0)$ are the initial conditions
associated with the trajectory which passes through
$({\bf z},\zeta)$ at time $t$, and the integral
$\int_0^t\zeta\,dt^\prime$ is performed along this
trajectory.
Then, repeating the steps
leading to Eq.\ref{eq:ratio}, we again get
$\overline{\exp -\beta W} = \exp -\beta\Delta F$,
where the overbar now denotes an average over our
ensemble of NH trajectories.
Thus, Eq.\ref{eq:iden} remains valid (given the 
canonical distribution of
initial conditions specified by Eq.\ref{eq:nhic})
when the system is coupled to a 
Nos\' e-Hoover thermostat, 
as per Eqs.\ref{eq:nh1} and \ref{eq:nh2}.
This result is identically true:
no weak coupling assumption is necessary, 
nor do we need to assume that the evolution
is chaotic.

It may similarly be established that Eq.\ref{eq:iden}
is valid, without additional assumptions,
when the thermostat is numerically implemented
using the Metropolis Monte Carlo algorithm,
rather than Nos\' e-Hoover dynamics.
In that situation, both the system and the 
Hamiltonian evolve by discrete steps, 
and the work performed is
a sum of changes in $H_\lambda$, evaluated
at successive locations of the system 
in phase space\cite{hunter}.

Numerical simulations of this sort are often used
to compute free energy differences of physical,
chemical or biological interest\cite{reviews}.
Typically, a number of simulations of slow
switching from one configuration to another
are performed, and the resulting average work
is used as an upper bound on $\Delta F$, as per
Eq.\ref{eq:average};
reversing direction, a lower bound is 
established \cite{hunter}.
The central result of the present paper
may be useful in this situation:
rather than taking the straight average of $W$, 
one can instead perform the average of $\exp -\beta W$,
then take the logarithm and multiply by 
$-\beta^{-1}$, as per Eq.\ref{eq:iden_b}.
In principle this converges to the exact value of
$\Delta F$ (rather than to an upper or lower bound)
as the number of simulations tends to infinity.
In practice, however, the same note of caution
applies here as in the case of coupling to a 
physical heat bath:  
if the fluctuations in $W$ from one simulation
to the next are much larger than $k_BT$, then
prohibitively many simulations may be necessary to
determine $\overline{\exp-\beta W}$ with
the desired accuracy.
Thus, we may expect Eq.\ref{eq:iden} to be
useful in free energy computations, only if
$\sigma$ is not much larger than $\beta^{-1}$.
Whether or not this condition holds for a given
system will depend on factors such as the number
of degrees of freedom, the switching time $t_s$, 
the switching path $\gamma$, and the numerical 
implementation of the heat bath. 

To summarize, the central result of this paper
is an equality which gives the
free energy difference $\Delta F$ between 
two configurations $A$ and $B$ of a 
classical, parameter-dependent system, 
in terms of an ensemble of finite-time measurements 
of the work performed on the system as it is 
switched from $A$ to $B$.
The derivation of this result relies on the 
assumption of weak coupling between system 
and reservoir, but otherwise follows directly 
from the properties of Hamilton's equations.
Two well-known equilibrium identities for
$\Delta F$, Eqs.\ref{eq:adia} and \ref{eq:dia},
emerge as limiting cases of this more
general, non-equilibrium result.
Practical considerations in all likelihood
limit the applicability of Eq.\ref{eq:iden}
to systems of no more than a moderate number of
degrees of freedom (e.g., nanoscale systems).
Finally, the equality may be useful when
numerical simulations of thermostatted 
systems are used to compute free energy 
differences.

\section*{ACKNOWLEDGMENTS}

It is a pleasure to acknowledge that numerous
stimulating discussions --- including those
with G.F.Bertsch, A.Bulgac, M.DenNijs,
P.DeVries, G.J.Hogenson, J.Hunter III, 
D.B.Kaplan, W.P.Reinhardt, T.Schaefer, D.Thouless,
and R.Venugopalan --- contributed to the 
derivation and understanding of the results
presented in this paper.
This work was supported by the Department of 
Energy under Grant No. DE-FG06-90ER40561.


\begin{references}
\bibitem{walls}
An exception to this statement occurs if
the system in question is
confined within an box with perfectly
hard walls, and if changing $\lambda$ corresponds
to moving these walls. 
In this case, Eq.\ref{eq:iden} remains true
for any finite $t_s$, but 
does not reduce to Eq.\ref{eq:dia} in the
limit $t_s\rightarrow 0$.
\end{references}
\end{document}